\documentclass[twocolumn,showpacs,preprintnumbers,amsmath,amssymb,floatfix]{revtex4}

\usepackage{epsfig,psfrag}
\usepackage{dcolumn}
\usepackage{bm}
\usepackage{graphicx}
\usepackage{color}

\begin{document}

\title{Nonlinear magnetotransport in interacting chiral nanotubes}
\author{A. De Martino,$^{1}$ R. Egger,$^1$ and A.M. Tsvelik$^2$}    
\affiliation{
${}^1$~Institut f\"ur Theoretische Physik, Heinrich-Heine-Universit\"at,
D-40225  D\"usseldorf, Germany \\
${}^2$~Department of Condensed Matter Physics and Materials Science,
Brookhaven National Laboratory, Upton, New York 11973-5000, USA}

\date{\today}

\begin{abstract}
Nonlinear transport through interacting single-wall nanotubes
containing a few impurities is studied theoretically.  
Extending the Luttinger liquid theory to incorporate 
trigonal warping and chirality effects, 
we derive the current contribution 
$I_e$ {\sl even}\ in the applied voltage $V$ and {\sl odd}\ in 
an orbital magnetic field $B$, which is non-zero only
 for chiral tubes and in the presence of interactions.
\end{abstract}
\pacs{71.10.Pm, 72.20.My, 73.63.Fg}

\maketitle
Electronic transport experiments on
individual single-wall carbon nanotubes (SWNTs)
\cite{bockrath}
have revealed ample evidence for the Luttinger liquid (LL)
phase of one-dimensional (1D) interacting metals
induced by electron-electron (e-e) interactions.
In its simplest form, the effective low-energy theory
for interacting SWNTs \cite{egger97} 
is insensitive to the chiral angle $\theta$ \cite{ando,chiralfoot}
describing the wrapping of the graphene sheet.
This fact can be rationalized by noting that, to lowest order
in a ${\bf k \cdot  p}$ scheme, the graphene dispersion 
reflects an isotropic Dirac cone around each K point
in the first Brillouin zone
\cite{ando}. Imposing periodic boundary
conditions around the SWNT circumference slices this cone 
and gives identical dispersion for all $\theta$,
as long as the SWNT stays metallic.
Such an approach is however insufficient for a description 
of the magnetotransport effects in chiral tubes. 
Therefore, we extend the theory \cite{egger97}
to include chirality effects by taking into account 
trigonal warping, tube curvature, and magnetic field $B$,
and then compute the
{\sl nonlinear} two-terminal magnetoconductance.
While the well-known Onsager symmetry 
 $G(B)=G(-B)$ \cite{onsager} excludes linear-in-$B$ terms in
the linear conductance, such terms can appear
out of equilibrium  \cite{rikken,spivak02,mesosc},
with first experimental observations reported
for SWNTs \cite{krstic,cobden} and semiconductor quantum dots or rings
 \cite{lofgren,marlow,ensslin,zumbuhl,helene}.

The current contribution $I_e$  {\sl odd}\ in $B$  
and {\sl even}\ in the voltage $V$ is of fundamental 
and unique importance, mainly due to two reasons. 
First, it requires
 a {\sl non-centrosymmetric (chiral) medium}, with the sign of $I_e$ 
depending on the handedness (enantioselectivity)
 \cite{rikken}, since the current density 
is a polar vector but magnetic field an axial one. Thus
$I_e\ne 0$ requires the simultaneous breaking of time reversal 
symmetry (by the magnetic field) and of inversion symmetry 
(by the chiral medium). Second, standard arguments based on 
the Landauer-B\"uttiker scattering formalism valid in the noninteracting 
case \cite{spivak02,mesosc} show that $I_e\ne 0$ also 
requires {\sl interactions}. At low temperature ($T$), 
e-e interactions should therefore contribute to $I_e$ in
leading order.  Measurements of $I_e$ probe interactions
and chirality in a very direct manner, potentially
allowing for the structural characterization of 
chiral $(\sin6\theta \ne 0)$ interacting nanotubes via
transport experiments.  Nonetheless, apart from the classical 
phonon-dominated high-$T$ regime \cite{spivak02},
no predictions specific to SWNTs have been made so far.
Here we determine the current contribution $I_e(V,T)$ 
linear in the parallel orbital magnetic
field $B$ (Zeeman fields play no role here), for
SWNTs in good contact to external leads.
The simplest case allowing for $I_e\neq  0$ is found 
when including at least two (weak) elastic scatterers, 
representing either defects or residual backscattering 
induced by the two contacts.  Our theory includes 
the often strong e-e interactions using
the bosonization method \cite{gogolin},
and holds for arbitrary chiral angle $\theta$.
Interactions are characterized by the LL parameter $K\leq 1$
in the charge sector, see Eq.~(\ref{LLH}) below,
with typical estimate $K\approx 0.2$ \cite{egger97}.
Our result for $I_e$, see Eqs.~(\ref{result}) and (\ref{spinless}), 
is non-zero only for chiral ($\sin 6\theta\ne 0$) 
interacting ($K<1$) SWNTs, and
changes sign for different handedness ($\theta\to -\theta$).
We predict oscillatory behavior of $I_e$ as a function
of bias voltage, where the oscillation period  
depends on the Luttinger parameter $K$, see Eq.~(\ref{dv}).
The coefficient \cite{cobden}
\begin{equation}\label{alphaT}
\alpha(T) = \lim_{V\to 0} \frac{I_e(V,T)}{V^2 B}
\end{equation}
is shown to exhibit power-law scaling at $T\to 0$, with a 
negative exponent that is again determined by $K$.  
Including only elastic impurity backscattering, $\alpha(T)$  does not
change sign with $T$ for otherwise fixed parameters.

Chirality effects in the low-energy theory
come about when one includes trigonal warping in the band structure. 
Its main effect is to introduce different Fermi 
velocities $v_{r\alpha}$ \cite{footmom}
for right- and left-moving excitations ($r=R/L=\pm$) 
around the two distinct K points ($\alpha=\pm$). All
these velocities coincide when disregarding trigonal warping,
and in a nominally metallic SWNT are given by
$v_{r\alpha}=v=8\times 10^5$~m$/$sec.  
For clarity, we consider an electron-doped SWNT
with equilibrium chemical potential $\mu\equiv \hbar v k_F>0$ 
well inside the conduction band, and omit the valence band.
Including the trigonal warping to lowest nontrivial order
\cite{ando,spivak02}, we find
\begin{eqnarray}\nonumber
v_{r\alpha} &=& v \sqrt{1-\frac{k_{\perp\alpha}^2}{k_F^2}} \Biggl ( 
1 +  \alpha \cos{3\theta}\frac{k_{\perp\alpha} a}{4\sqrt{3}}
\frac{3-2k_{\perp\alpha}^2/k_F^2}{
1-k_{\perp\alpha}^2/k_F^2} \\   & & \label{velocities}
-r \alpha \sin{3\theta}
\frac{k_F a}{2\sqrt{3}} \sqrt{1-\frac{k_{\perp\alpha}^2}{k_F^2}}\Biggr).
\end{eqnarray}
The effects of tube curvature and magnetic field enter via the quantized
transverse momenta
\begin{equation}\label{transmom}
k_{\perp\alpha} R = n-\alpha\nu/3 + \Phi/\Phi_0 + \alpha c \cos(3\theta),
\end{equation}
where $a=0.246$~nm is the lattice spacing, 
$R$ the tube radius, $\Phi=\pi R^2 B$, $\Phi_0= h/e$ the flux quantum,
and the tube curvature results in $c=\kappa a/R$ with 
$\kappa\approx 1$ \cite{ando}. 
Here integer $n$ with $|k_{\perp\alpha}|<k_F$ are allowed; 
we focus on the lowest band $n=0$ in what follows.  
The index $\nu$ distinguishes nominally metallic ($\nu=0$) 
and semiconducting ($\nu=\pm 1$)
SWNTs, and for simplicity, from now on we assume $\nu=0$. 
However, with minor modifications, our theory below also 
applies to strongly doped semiconducting tubes.
We mention in passing that chirality effects are also important
for other physical quantities.  In particular, scattering by a 
long-range potential in metallic SWNTs 
depends on chirality
\cite{kane1,mceuen}.

In terms of annihilation fermion operators 
$R_{\alpha\sigma}(x)$ and $L_{\alpha\sigma}(x)$ for 
right- and left-movers of spin $\sigma=\pm$, respectively,
the usual linearization of the band structure around the Fermi points
then yields $H=H_{LL}+V_{dis}$ (we set $\hbar=k_B=1$), 
\begin{eqnarray}
H_{LL} & =& -i \sum_{\alpha\sigma}\int d x
\left( v_{R\alpha} R^\dagger_{\alpha\sigma}
 \partial_x R_{\alpha\sigma}^{} - v_{L\alpha}
 L^\dagger_{\alpha\sigma}\partial_x L^{}_{\alpha\sigma}\right) \nonumber\\
& & + \frac{V_0}{2} \int dx \left(
 R^\dagger R + L^\dagger L \right)^2 ,\label{ll} \\ 
V_{dis} & =& \int d x \ U(x) \left( 
e^{-2i k_F x} R^\dagger L + {\rm h.c.} \right) , \label{dis}
\end{eqnarray}
where $4V_0/\pi v =K^{-2}-1$ describes e-e
forward scattering interactions \cite{foot},
and $\alpha\sigma$ summations are implied
when not given explicitly.  Elastic disorder leads to 
$V_{dis}$, e.g., due to impurities, defects or substrate 
inhomogeneities. We keep only the intra-band backscattering 
potential $U(x)$, which yields the dominant 
impurity effect \cite{egger97}. 
This Hamiltonian can be efficiently treated by
(Abelian) bosonization \cite{gogolin}.
Introducing four bosonic fields $\phi_i(x)$ and their 
dual $\theta_i(x)$, with $i=(c+,c-,s+,s-)$ 
denoting the total/relative charge/spin modes,
the clean Hamiltonian for $v_{r\alpha}=v$ is \cite{egger97}
\begin{eqnarray}
H^{(0)}_{LL} &=& \frac{u_{c}}{2} \int dx \left[ 
K^{-1} (\partial_x \phi_{c+})^2 + K (\partial_x \theta_{c+})^2 \right] 
\nonumber \\ 
&+& \frac{u_n}{2} \sum_{i\ne c+} \int dx \left[ (\partial_x \phi_i)^2 + 
(\partial_x \theta_i)^2 \right] , \label{LLH}
\end{eqnarray}
where $u_{c}=v/K$ is the plasmon velocity for the
$c+$ mode, and $u_n=v$ for the three neutral modes.
Including the $v_{r\alpha}$ differences in Eq.~(\ref{velocities})
then brings about two new features, acting separately   
in the decoupled charge ($i=c\pm$) and spin ($i=s\pm$) sectors:
(i) couplings between total ($+$) and relative ($-$) modes, and (ii) 
couplings between mutually dual ($\theta_i,\phi_i)$ fields. The first
leads to tiny quantitative corrections but no 
qualitative changes and is neglected henceforth.
Point (ii) is crucial, however, since it 
implies different velocities ($u_{c,R/L}$ and $u_{n,R/L}$) for 
right- and left-moving plasmons. This eventually 
produces the current contribution $I_e$ in the presence 
of impurities. To linear order in $B$, we find
\begin{eqnarray}
\label{veldiff}
u_{c,R/L}/v &=& 1/K\pm \delta, \quad u_{n,R/L}/v =1\pm \delta , \\
\nonumber
\delta &=& \frac{\Phi/\Phi_0}{2\sqrt{3} k_F R} (a/R)^2 \ \sin( 6 \theta) .  
\end{eqnarray}
Note that $\delta\ne 0$ requires both $B\ne 0$ and chirality, 
$\sin 6\theta \ne 0$.
Moreover, $\delta$ has opposite sign for opposite handedness.
It essentially describes the difference of $R/L$-moving velocities, 
and depends linearly on $B$.

Next we address the computation of $I_e(V,T)$, 
where the current operator 
is $(2e/\sqrt{\pi}) \partial_t \phi_{c+}$.
An important point concerns the inclusion of
the applied voltage $V$ in a two-terminal setup. 
For weak impurity backscattering $U(x)$,
it is sufficient to address the clean case with 
adiabatically connected leads, where 
a time-dependent shift arises \cite{eggergrabert},
$\phi_{c+}\to \phi_{c+}+eVt/\sqrt{\pi}$.
However, with chiral asymmetry, $v_R\ne v_L$,
there is an additional effect  \cite{epl} 
due to the different density of states $\nu_{R/L}=2/\pi v_{R/L}$
for $R/L$ movers.  Starting from the equilibrium chemical potential $\mu$, 
when applying a voltage, the chemical potentials $\mu_{R/L}$
of $R/L$ movers are set by the left/right reservoirs,
respectively, where $\mu_R-\mu_L=eV$.
Relative to the equilibrium density, $R/L$ movers are
thereby injected with density 
$\rho^0_{R/L}= \nu_{R/L} (\mu_{R/L}-\mu)$.
We may now write $\mu_{R/L} = \mu +\Delta \mu \pm eV/2$, 
where $\Delta \mu$ is determined by the condition that
in an adiabatically connected impurity-free quantum wire, no
charge can accumulate in the steady state \cite{eggergrabert},
$\rho^0_R+\rho^0_L=0$.  This yields $\Delta \mu=\delta eV/2$, 
where $\delta=(v_R-v_L)/(v_R+v_L)$ has been specified
in Eq.~(\ref{veldiff}), implying the voltage-dependent shift   
\begin{equation}\label{sh2}
k_F \to k_F + \delta e V/2 v_0, \quad 
v_0= \frac{2v_Rv_L}{v_R+v_L}=v(1-\delta^2)
\end{equation}
in Eq.~(\ref{dis}).  Under the non-equilibrium Keldysh formalism,
to second order in $U(x)$, $I_e$ then follows as
\begin{eqnarray} \label{result}
I_e &=& - \frac{2eK}{(\pi a)^2} 
\int_{-\infty}^\infty dt \int dX dx \sin(2k_F x) 
\\ \nonumber &\times& U(X+x/2)\,  U(X-x/2) 
\\ \nonumber  &\times&  \cos(eV [t + 
\delta x/v_0]) \
{\rm Im}  e^{i \pi {\cal G}^<(x,t)},
\end{eqnarray}
where ${\cal G}^<$ is the sum of the lesser Green's functions for the
four plasmonic modes, see Eq.~(\ref{corr}) below. 

As a function of time,  $e^{i{\pi \cal G}^<(x,t)}$ has singularities  
only in the lower-half complex plane, and therefore $I_e=0$ for $V=0$.
By inspection of Eq.~(\ref{result}), 
we also observe that $I_e(-\delta)= - I_e(\delta)$. With Eq.~(\ref{veldiff})
we conclude that $I_e$ is {\em odd} both in magnetic field and chiral angle. 
In fact, $I_e$ is the only odd-in-$B$ contribution, since 
the odd-in-$V$ part of the current turns out to be {\sl even} in $B$.
We can now state a first necessary condition for $I_e\neq 0$, namely
$\delta\neq 0$,
which requires the simultaneous breaking of time reversal and 
inversion symmetry.
Approximating the impurity potential as
 $U(x) = \sum_i U_i\delta(x - x_i)$, Eq.~(\ref{result}) becomes
\begin{eqnarray} \label{result1}
I_e&=& - \frac{4eK}{(\pi a)^2} \sum_{j>k}U_j\, U_k \sin (2k_F x_{jk}) 
\\ \nonumber
&\times&\int d t \cos(eV [t +\delta  x_{jk}/v_0]) 
\ {\rm Im} \ e^{i\pi {\cal G}^<(x_{jk},t)} 
\end{eqnarray}
with $x_{jk}=x_j-x_k$.  Then a second necessary condition arises:
there should be just a few impurities.  The function ${\cal G}^<(x,t)$ 
changes slowly on the scale $k_F^{-1}$, and 
when one has to sum over many impurities,
$I_e=0$ due to the fast oscillations of $\sin(2k_F x_{jk})$ 
in Eq.~(\ref{result1}).  In such a disordered case, however,
{\sl fluctuations} exhibit related magnetochiral 
transport effects \cite{mesosc}.
When only a few impurities (but at least two) are present,  
there is no averaging and the effect survives in the current, 
albeit its magnitude and sign are of course sample-dependent.
It is likely that the experiments of Ref.~\cite{cobden} were 
performed on samples with
not too many impurities, for otherwise  strong 
localization effects characteristic
for 1D systems would render them insulating.
Below we focus on the simplest case of two 
impurities separated by a distance $L=x_2-x_1$, see also 
Ref.~\cite{sanchez}.
Such a controlled two-impurity setup can be
experimentally realized in individual SWNTs \cite{postma}.

As further evaluation of Eq.~(\ref{result1}) requires numerical
integration routines, we first examine the spinless single-band version 
of Eq.~(\ref{ll}), which contains the
essential physics and allows to compute $I_e$ in closed form
\cite{footform}.  Given the Fermi velocities $v_{r=R/L=\pm}=v(1\pm \delta)$ 
[corresponding to Eq.~(\ref{velocities})],
the plasmon velocities of the clean interacting system
 are $u_{R/L}=v(1/K\pm \delta)$ [corresponding to 
Eq.~(\ref{veldiff})].  The plasmon lesser Green's 
function is
\begin{equation}\label{corr}
{G}^<(x,t) = \frac{iK}{4\pi}\sum_r \ln \left[
\frac{u_r}{\pi a T} \sinh\left( \frac{ \pi T}{u_r} [x-ira -
r u_r t]\right)
\right].
\end{equation}
Using $2u_R u_L/(u_R+u_L)\simeq v/K$, some algebra yields 
\begin{eqnarray}\nonumber
I_e &=& \frac{2eK^2}{\pi L} \frac{U_1 U_2}{v} 
(a/L)^{2K-2} \sin(2k_F L)\ \tilde{T}^{2K-1} e^{- K \tilde T} \\
\label{spinless}
&\times& \sin[\delta(1-K^2)\tilde V/K] \ {\rm Im} \Biggl[ 
\frac{\Gamma(1+K-i\tilde V/\tilde T)}
{\Gamma(K)\Gamma(2-i\tilde V/\tilde T )} \\
\nonumber &\times& e^{i\tilde V}
F \left(K,1+K-i\tilde V/\tilde T; 2-i\tilde V/\tilde T; e^{-2\tilde{T}} 
\right) \Biggr ],
\end{eqnarray}
where $\Gamma$ is the Gamma function and $F$ the hypergeometric
function \cite{gradsteyn}. 
We introduced dimensionless temperature and voltage
\begin{equation}\label{ttilde}
\tilde T \equiv  \frac{ 2\pi k_B T} {\hbar v /(K L)},
\quad \tilde V \equiv \frac{|eV|}{\hbar v/(KL)} .
\end{equation}
Obviously, $I_e=0$ in the noninteracting case ($K=1$),
and thus interactions provide a third necessary condition for $I_e\ne 0$.
Equation (\ref{spinless}) now allows to analyze several limits of interest.
First, one recognizes an oscillatory dependence on the doping
level $\mu=v k_F$ tuned by a backgate voltage, similar
to what is seen experimentally \cite{cobden}. 
Second, $\alpha(T)$ [see Eq.~(\ref{alphaT})]  is 
exponentially small for $\tilde T\gg 1$ but 
shows power-law scaling $\alpha(\tilde T \ll 1) \propto \tilde T^{2K-2}$  
at low temperatures, implying a huge 
increase in $\alpha(T)$ when
lowering $T$.  Third, in the zero-temperature limit, 
Eq.~(\ref{spinless}) yields 
\begin{equation} \label{zerot}
I_e(\tilde V)\propto \sin[\delta(1-K^2)\tilde V/K] \
\tilde V^{K-1/2} J_{K-1/2}(\tilde V)
\end{equation}
with the Bessel function $J_\nu$ \cite{gradsteyn}. This implies
the low-voltage scaling $I_e\propto \tilde V^{2K}$, {\sl a posteriori}\ 
justifying our perturbative treatment of the impurity potential.
Remarkably, $I_e$ shows {\sl oscillatory behavior}\ as a function
of the bias voltage $V$. As one can see from
Eq.~(\ref{zerot}),  there are two different oscillation
periods,
\begin{equation}\label{dv}
\Delta V_1 = \frac{2\pi \hbar v } { e K L} ,\quad 
\Delta V_2 = \frac{ K \Delta V_1}{\delta (1-K^2)}.
\end{equation}
For strong interactions, $\Delta V_2$ can in principle
provide direct information about $\delta$.
In any case, $\Delta V_1$ should be readily observable 
and yields already the Luttinger parameter $K$.
Inspection of Eq.~(\ref{result1}) suggests that the
physical reason for these oscillations is the
quantum interference between right- and left-moving
waves travelling between the two impurities with
different velocities.  

\begin{figure}[t!]
\scalebox{0.28}{\includegraphics{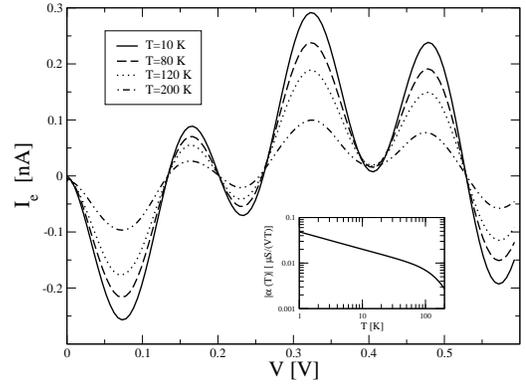}} 
\caption{ \label{fig1} 
Even-in-voltage part of the current $I_e$ (in nA) 
as a function of bias voltage (in V) for several
temperatures.  Results are shown for a $(10,4)$ SWNT
with $K=0.23$, $B=16$~T, containing two impurities ($L=20$~nm,
$U_{1,2}=v/2$). Inset: Coefficient $\alpha(T)$ [in $\mu S V^{-1} T^{-1}$] 
defined in Eq.~(\ref{alphaT}) as a function of 
temperature in double-logarithmic scales.
}
\end{figure}

Let us then go back to the full four-channel case, and perform 
the integration in Eq.~(\ref{result1}) numerically.
To be specific, we consider a long and
adiabatically connected $(10,4)$ SWNT with two symmetric 
impurities separated by $L=20$ nm. Fig.~\ref{fig1}
shows $I_e$ as a function of $V$ for several $T$.
Oscillatory behaviors with bias voltage are clearly visible.
With decreasing temperature, $I_e$ increases and 
one gets power-law scaling of $\alpha(T)\propto T^{(K-1)/2}$ 
at low $T$ (see inset of Fig.~\ref{fig1}), 
generalizing the above single-channel result.
Notably, $\alpha(T)$ has the same
sign for a given parameter set, in qualitative 
agreement with experiments \cite{cobden}.  
Within our parameter choices, also the order
of magnitude in $\alpha(T)$ agrees with Ref.~\cite{cobden}.
In any case, the picture obtained in the single-channel 
version is essentially recovered under the four-channel calculation.

To conclude, we have analyzed nonlinear magnetochiral transport 
properties of interacting single-wall carbon nanotubes.  In
chiral tubes, measurement of the  odd-in-$B$ component $I_e$
(which must be even in $V$)
provides direct information about interactions and
chirality not accessible otherwise.  For two impurities, we
have presented detailed analytical results for $I_e$. We predict
oscillations of $I_e$ as a function of 
bias voltage, which provide direct information about the interaction parameter
$K$. Moreover, at low temperatures, power-law scaling is found and leads to an 
enhancement of $I_e$. 
In future work, our approach should be useful when calculating
fluctuations of transport coefficients in disordered
interacting SWNTs.

We thank D. Cobden, F. Essler, E. Papa, A. Shytov, and B. Spivak 
for discussions.
ADM and RE are grateful to the Institute for Strongly Correlated 
and Complex Systems at BNL for hospitality during their extended visits.
This work was supported by the DFG-SFB TR 12 (RE), the ESF network 
INSTANS (ADM and RE), and 
the US DOE under contract number DE-AC02-98 CH 10886 (AMT).


\begin{thebibliography}{99}
\bibitem{chiralfoot}
For a $(n,m)$ SWNT, the chiral angle is $\theta=\tan^{-1}[\sqrt{3}m/(2n+m)]$.
The $(n,m)$ indices determine whether the SWNT is metallic or semiconducting
\cite{ando}.
\bibitem{footmom}
Fermi momenta are also slightly different for 
$R/L$ movers and/or in different bands.
For a calculation of $I_e$, this is inessential and neglected henceforth.
\bibitem{foot} Electron-electron backscattering effects
in SWNTs  are tiny \cite{egger97} and disregarded here.
\end{thebibliography}
\end{document}